\documentclass[reqno,12pt]{amsart}
\usepackage{amsmath,amsfonts,amssymb}
\usepackage{graphicx}
\usepackage{color}
\usepackage{float}
\usepackage{color}   

\def\beq#1#2\eeq{\begin{equation}\label{#1}#2\end{equation}}
\def\bal#1#2\eal{\begin{align}\label{#1}#2\end{align}}
\def\bse#1#2\ese{\begin{subequations}\label{#1}#2\end{subequations}}

\def\dd{\operatorname{d}}

\begin{document} 
\def\singlespacing{\baselineskip=13pt}	\def\doublespacing{\baselineskip=18pt}
\doublespacing

\pagestyle{myheadings}\markright{{\sc  Impedance matrix for  anisotropic cylinders   }  ~~~~~~\today}

\title[Stable impedance methods]{
{Stable methods to solve the impedance matrix for  radially inhomogeneous cylindrically anisotropic structures}  
}

\author[Norris et al.]{ Andrew N. Norris, Adam J. Nagy, \& Feruza A. Amirkulova}
\address{Mechanical and Aerospace Engineering, Rutgers University, Piscataway NJ 08854}

\date{\today}

\begin{abstract}
A stable approach for  integrating the impedance matrix in cylindrical, radial inhomogeneous structures is developed and studied.  A Stroh-like system using the time-harmonic displacement-traction state vector is used to derive the Riccati matrix differential equation involving the impedance matrix.  It is found that the resulting equation is stiff and leads to exponential instabilities.  A stable scheme for integration is found in which a  local expansion is performed by combining the matricant and impedance matrices.  This method offers 
a stable solution for fully anisotropic materials, which was previously unavailable in the literature.  
Several approximation schemes for integrating the Riccati equation in cylindrical coordinates are considered: exponential, Magnus, Taylor series, Lagrange polynomials, with numerical examples indicating that  the exponential scheme  performs best. 
 The impedance matrix is compared with solutions involving Buchwald potentials in which the material is limited to piecewise constant transverse isotropy.  Lastly a scattering example is considered and compared with the literature.

\end{abstract}

\maketitle



%
\section{Introduction}\label{sec1}

Wave propagation in layered elastic media has been widely  studied resulting in a variety of solution approaches.  These include the use of  scalar and vector potentials \cite{MorseI},  the transfer matrix method \cite{Brekh60,Huang95,Sinclair93,Rokhlin96,Honarvar97}, and the delta matrix method   \cite{Thrower65,Dunkin65}. Alternative, computationally stable methods have also been developed, e.g. the stiffness matrix  \cite{Rokhlin2002,Rokhlin04}, the global matrix \cite{Schmidt85}, and the reflectivity method \cite{Rokhlin92}. Such approaches are limited to isotropic or transversely isotropic materials whereas we are interested in general anisotropic solids in order to develop scattering solutions related to metamaterial devices such as acoustic cloaks \cite{Norris11a,Norris08b,Scandrett10} which can be modeled as radially inhomogeneous anisotropic solids.  The goal of this paper is to produce a stable solution method for such materials.

We considered a matricant propagator solution \cite{Shuvalov03,Norris10} based on Stroh formalism to solve for scattering from  a generally anisotropic material.  The method involves the creation of a state space representation of the system where six first order, ordinary differential equations, must be integrated which may often diverge.  A stable scheme for finding the solution to the differential equations, inspired by \cite{Schiff99}, is developed by combining the matricant and impedance matrices.  The scheme is considered with several different expansion methods to yield relatively high orders of accuracy and is compared with solutions from the conditional impedance matrix and Buchwald's scalar potentials \cite{Buchwald}.  

The outline of the paper is as follows.  In section \ref{sec2}, we begin with definitions  of the 
impedance and matricant matrices.   
 An explicit method for finding the impedance in piecewise uniform, transversely isotropic materials is developed in section \ref{sec3}.  This method also  serves as  a tool to compare with more general solution methods based on the 
Riccati matrix differential equation for the impedance matrix. 
 It is found that the Riccati  differential equation is stiff and leads to exponentially growing instabilities.  In section \ref{sec4}, an alternative approach for integrating the matricant is derived to find a stable means of integrating the impedance matrix.  We then use different expansion techniques used in the integration process in order to find higher order accurate schemes.  Lastly in section \ref{sec5} a scattering example from the literature is considered.  


\section{Impedance and matricant matrices}\label{sec2}
 
We consider time harmonic wave motion in radially inhomogeneous cylindrically anisotropic solids.   The associated 
equilibrium equations for linear elastodynamics in cylindrical coordinates are summarized in \ref{appA}.   Here we need only focus on the relation between the  vectors $\mathbf{U}(r)$ and $  \mathbf{V}(r)$ associated with displacement and traction, respectively.  Precise definitions follow from Eqs.\ \eqref{-1-11} (which includes the superscript $n$ that is here omitted for simplicity).  The dimension of each vector is taken as $m$, where $m$ is either $3$, $2$ or $1$; $m=3$ in general, $m=2$ if    $z-$dependence is not considered, and $m=1$ for pure out-of-plane
shear horizontal (SH) motion.  For the moment we may consider $m$ as general.  The main focus of the paper is the  $m \times m$  conditional  impedance matrix ${\bf z}$ defined 
  such that  
\beq{00}
\mathbf{V} (r) =- \text{i} {\bf z} (r) {\mathbf U}(r) .
\eeq

It is shown in Eq.\ \eqref{-1-12}  that the equations for linear elastodynamics can be cast as a
 system of $2m$ linear ordinary differential equations  \cite{Norris10} 
\begin{equation}
 \frac{\dd \boldsymbol{\eta} }{\dd r} =
\mathbf{Q}\boldsymbol{\eta} \ \quad \mathrm{with}
\ \boldsymbol{\eta} (r) = \begin{pmatrix}
\mathbf{U}  \\ 
\mathbf{V}
\end{pmatrix},
\ 
\ \mathbf{Q}(r) = 
\begin{pmatrix}
\mathbf{Q}_{1} & \mathbf{Q}_{2} \\ 
\mathbf{Q}_{3} & \mathbf{Q}_{4}%
\end{pmatrix}
 ,  \label{01}
\end{equation}
where $\mathbf{Q}^+ = - \mathbf{T}\mathbf{Q} \mathbf{T} $, $^+$ denotes the Hermitian transpose  and 
  $\mathbf{T}$ is defined in \eqref{-1-19}. 
It follows from Eqs.\ \eqref{00} and \eqref{01} that  ${\bf z} (r)$ satisfies a differential Riccati equation 
\begin{equation}
\frac{\dd {\bf z}}{\dd r} +{\bf z} {\mathbf Q}_{1}-\mathbf{Q}_{4}{\bf z} -i{\bf z} \mathbf{Q}%
_{2}{\bf z}-i\mathbf{Q}_{3}=\mathbf{0},  \label{2}
\end{equation}%
with assumed initial condition ${\bf z} (r_0)$  at some specified $r=r_0$, hence the name  conditional impedance. 

One approach to solving for the conditional  impedance matrix, ${\bf z}$, is to first 
solve for the $2 m \times 2 m$  matricant $\mathbf{M}  ( r,r_0 )$ which is   defined as 
the solution of the initial value problem 
\begin{equation}\label{32}
  \frac{\dd \mathbf{M} }{\dd r}  
  ( r,r_0 )=\mathbf{Q}(r) \mathbf{M} ( r,r_0 ),
	\ \mathbf{M} ( r_0,r_0) = \mathbf{I}_{(2m )}, 
	\ 
\mathbf{M} =  \begin{pmatrix}
\mathbf{M}_{1} & \mathbf{M}_{2} \\ 
\mathbf{M}_{3} & \mathbf{M}_{4}%
\end{pmatrix}
.
\end{equation}
Hence 
\beq{4=67}
\boldsymbol{\eta} (r) = \mathbf{M} ( r,r_0 )\boldsymbol{\eta} (r_0). 
\eeq
 Using   the relations 
\beq{045}
\mathbf{U}\left( r \right) =\left( \mathbf{M}_{1}-\text{i}\mathbf{M}_{2}{\bf z} (r_0) \right) \mathbf{U}\left( r_0\right) ,\quad 
\mathbf{V}\left( r\right) =\left( \mathbf{M}_{3}-\text{i}\mathbf{M}_{4}{\bf z} (r_0) \right) \mathbf{U}\left( r_0\right) ,%
\eeq
which follow from  \eqref{00}, the conditional impedance  can be expressed in terms of the matricant as  
\beq{46}
{\bf z} (r) = i\big( \mathbf{M}_{3}-i\mathbf{M}_{4}{\bf z} (r_0)  \big) \big( \mathbf{M}_{1}-i\mathbf{M}_{2}{\bf z} (r_0) \big)^{-1} .
\eeq
The propagator nature of the matricant is apparent from Eq.\ \eqref{4=67} and from the property  
 $\mathbf{M}  ( r,r_1 )$ $ \mathbf{M}  ( r_1,r_0 )$ $= \mathbf{M}  ( r,r_0 )$, and in particular $\mathbf{M}  ( r,r_0 )
 = \mathbf{M}  ( r_0,r )^{-1}$.  Also,  the   symmetry \eqref{-1-19} implies 
 $
      \mathbf{M} (r,r_0) = \mathbf{T}\mathbf{M}^+( r_0,r) \mathbf{T}$. 
Hence,  $
      \mathbf{M}^{-1} (r,r_0) = \mathbf{T}\mathbf{M}^+( r,r_0) \mathbf{T} $, 
 that is, $\mathbf{M}$    is  $\mathbf{T}$-unitary \cite{Pease}. 

An alternative approach to finding ${\bf z}$ uses the two point impedance matrix, which by definition  relates the traction and displacement vectors at two values of $r$   according to  \cite{Norris10} 
\begin{equation}\label{i35}\begin{small}
\begin{pmatrix} \mathbf{V}(r_0) \\ -  \mathbf{V}(r) \end{pmatrix}= - \text{i} \mathbf{Z}(r,r_0)
\begin{pmatrix} \mathbf{U}(r_0) \\ \mathbf{U}(r) \end{pmatrix},
\quad
	\ 
\mathbf{Z} =  \begin{pmatrix}
\mathbf{Z}_{1} & \mathbf{Z}_{2} \\ 
\mathbf{Z}_{3} & \mathbf{Z}_{4} 
\end{pmatrix}.
\end{small}\end{equation}
The two point impedance matrix has the important property that it is Hermitian, $\mathbf{Z}=\mathbf{Z}^+$ \cite{Norris10}.
The relations between the matricant of \eqref{32}  and the impedance matrix of \eqref{i35} evaluated at    cylindrical surfaces $r$, $r_0$ are easily deduced   \cite{Norris10} 
\begin{equation}\label{i56}
\begin{aligned}
 {\mathbf{M}}(r,r_0)
 &=
 \begin{pmatrix}
-{\mathbf{Z}}_{2}^{-1} {\mathbf{Z}}_{1}
 & \text{i} {\mathbf{Z}}_{2}^{-1}
  \\ 
 \text{i} {\mathbf{Z}}_{3} - \text{i} {\mathbf{Z}}_{4} {\mathbf{Z}}_{2}^{-1} {\mathbf{Z}}_{1}
   & - {\mathbf{Z}}_{4}  {\mathbf{Z}}_{2}^{-1}
\end{pmatrix},
\\
{\mathbf{Z}}(r,r_0)
& =
 \begin{pmatrix}
-\text{i} {\mathbf{M}}_{2}^{-1} {\mathbf{M}}_{1}
 &  \text{i} {\mathbf{M}}_{2}^{-1}
  \\ 
\text{i}  {\mathbf{M}}_{4} {\mathbf{M}}_{2}^{-1} {\mathbf{M}}_{1}- {\mathbf{M}}_{3}
  &-\text{i} {\mathbf{M}}_{4} {\mathbf{M}}_{2}^{-1}
\end{pmatrix}.
\end{aligned}
\end{equation}
Introducing \eqref{i56} into \eqref{46}, we can relate the conditional impedance ${\bf z}(r)$ to the two point impedance matrix ${\mathbf{Z}}(r,r_0)$ according to 
\beq{i62}  
{\bf z} (r) =  {\mathbf{Z}}_{3} \big( {\mathbf{Z}}_{1}     - {\bf z}(r_0) \big)^{-1} {\mathbf{Z}}_{2} 
- {\mathbf{Z}}_{4} . 
\eeq

\section{Piecewise uniform  transverse isotropy}\label{sec3}

In this section  we develop an  approach suitable for piecewise uniform transversely isotropic cylinders by explicit calculation of the   global impedance matrix $\mathbf{Z}$ of \eqref{i35}, from which the conditional impedance can be found using \eqref{i62}.   The method is based on 
   a recursive algorithm proposed by Rokhlin et al. \cite{Rokhlin2002}  called the stiffness matrix method.  The 	analysis in \cite{Rokhlin2002} was restricted to multilayered media in Cartesian coordinates, whereas the present method is applicable to cylindrically layered media of transverse isotropy.  We will refer to Rokhlin and Wang  \cite{Rokhlin2002}  several times in this section to note the similarities and differences of the  approaches.



\begin{figure}[h]
	\centering
		\includegraphics[width=5in]{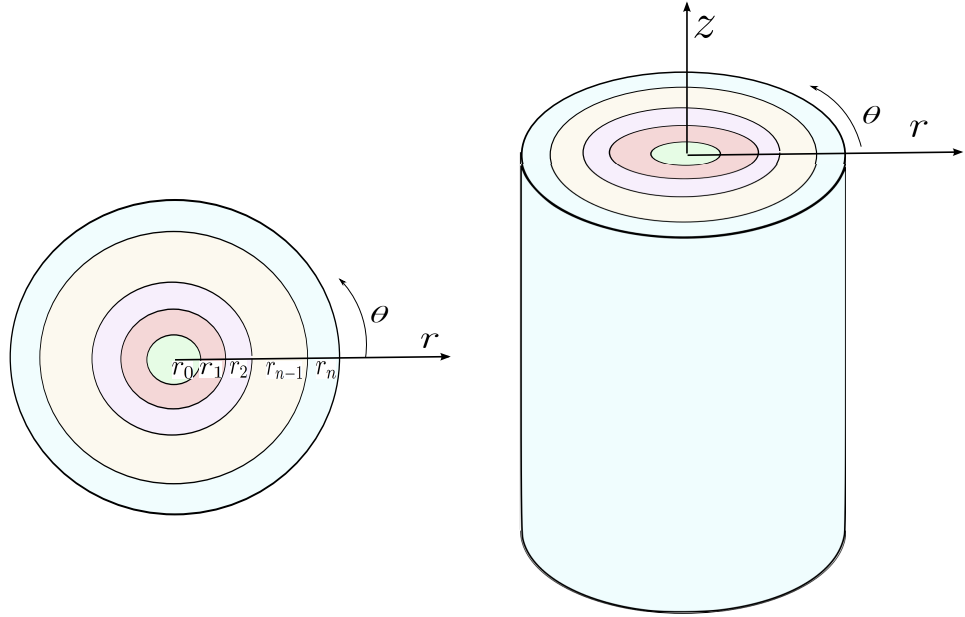}
		\caption{A cylindrically anisotropic multilayered media is considered in the  system of cylindrical coordinates. The media consists of $n$  anisotropic layers with different densities and elasticity tensors in general.}
	\label{fig0}
\end{figure}

Consider $n>1$  layers of uniform transversely isotropic materials with the $k\text{th}$ layer $r_{k-1} < r < r_k$, $k\in \overline{1,n}$, see Figure \ref{fig0}. The explicit form of the local two point impedance matrix  of the $k\text{th}$   layer   is 
$\mathbf{Z}^{k} (r_{k},r_{k-1})$ as defined by Eq.\ \eqref{i36}.  
Denote the  global two point impedance matrix for the   cylinder between $r_0$ and $r_k$ by 
$\mathbf{Z}^{\textbf{K}} = \mathbf{Z}^{\textbf{K}}(r_k,r_0)$.   Our objective is the 
global two point impedance matrix for the  entire  cylinder, 
$\mathbf{Z}(r_n,r_0) \equiv \mathbf{Z}^{\textbf{N}}(r_n,r_0)$.

Consider first the two bordering layers  between $r=r_0$ and $r=r_2$ and sharing the $r=r_1$ surface. 
Continuity of displacements and traction on the interface implies  
\bse{2=3}
\bal{i37}
\begin{pmatrix}  \mathbf{V}_{0} \\ - \mathbf{V}_{1} \end{pmatrix} 
&  
= -\text{i} \begin{pmatrix}
{\mathbf{Z}}_{1}^{a}  & {\mathbf{Z}}_{2}^{a} \\ 
{\mathbf{Z}}_{3}^{a} & {\mathbf{Z}}_{4}^{a}
\end{pmatrix}
\begin{pmatrix} \mathbf{U}_{0} \\ \mathbf{U}_{1} \end{pmatrix},
\\
\begin{pmatrix}  \mathbf{V}_{1} \\ - \mathbf{V}_{2} \end{pmatrix} 
& =
-\text{i} \begin{pmatrix}
{\mathbf{Z}}_{1}^{b}  & {\mathbf{Z}}_{2}^{b} \\ 
{\mathbf{Z}}_{3}^{b} & {\mathbf{Z}}_{4}^{b}
\end{pmatrix}
\begin{pmatrix} \mathbf{U}_{1} \\  \mathbf{U}_{2} \end{pmatrix} ,
\label{i38}
\eal
\ese
where  $\mathbf{Z}^{a}\equiv \mathbf{Z}^{1} (r_{1},r_0)   $,
$\mathbf{Z}^{b}\equiv \mathbf{Z}^{2} (r_{2},r_1)   $.
From the second row of Eq.\ \eqref{i37} and the first row of Eq.\ \eqref{i38}, we have 
\beq{i39}
\mathbf{U}_{1}= -\big({{\mathbf Z}}_{4}^{a}+{{\mathbf Z}}_{1}^{b} \big)^{-1}   \big(  {{\mathbf Z}}_{3}^{a} \,\mathbf{U}_{0} + {{\mathbf Z}}_{2}^{b}\,\mathbf{U}_{2} \big).
\eeq
Introducing   Eq.\ \eqref{i39} into  Eqs.\ \eqref{i37} and \eqref{i38}, we define the impedance matrix $ \mathbf{Z}^{\textbf{2}}(r_2,r_0) $ that relates the traction vector to the displacement vector on the inner $(r=r_0)$ and outer $(r=r_2)$ surfaces of the bilayer, 
\begin{equation}\label{i40}
\begin{pmatrix}  \mathbf{V}_{0} \\ - \mathbf{V}_{2} \end{pmatrix} = -  \text{i}\, \mathbf{Z}^{\textbf{2}}(r_2,r_0) \begin{pmatrix} {\bf U}_{0} \\  {\bf U}_{2} \end{pmatrix},
\end{equation}
where
\begin{equation}\label{i40X}
\begin{aligned}
 \mathbf{Z}^{\textbf{2}}(r_2,r_0)=
 \begin{pmatrix}
{{\mathbf Z}}_{1}^{a}-{{\mathbf Z}}_{2}^{a}\big({{\mathbf Z}}_{4}^{a} + {{\mathbf Z}}_{1}^{b} \big)^{-1} {{\mathbf Z}}_{3}^{a}
 & - {{\mathbf Z}}_{2}^{a}  \big({{\mathbf Z}}_{4}^{a} + {{\mathbf Z}}_{1}^{b} \big)^{-1}     { {\mathbf Z}}_{2}^{b}
  \\ 
 - {{\mathbf Z}}_{3}^{b}  \big({{\mathbf Z}}_{4}^{a}+{{\mathbf Z}}_1^{b} \big)^{-1}     { {\mathbf Z}}_{3}^{a}
  & {{\mathbf Z}}_{4}^{b} - {{\mathbf Z}}_{3}^{b} \big({{\mathbf Z}}_{4}^{a} + {{\mathbf Z}}_{1}^{b} \big)^{-1} {{\mathbf Z}}_{2}^{b}
\end{pmatrix},
\end{aligned}
\end{equation}
${\mathbf{Z}}_{i}^{a}={\mathbf{Z}}_{i}^{1}$, ${\mathbf{Z}}_{i}^{b}= {\mathbf{Z}}_{i}^{2}$ and ${\mathbf{Z}}_{i}^{k}\, (k=1, 2, \, i=\overline{1,4})$ are given by Eqs.\ \eqref{i36}. 
Note that Eqs.\ \eqref{i37} and \eqref{i38}  are similar, apart from a sign change, to Eqs.\ (19) and (20) in \cite{Rokhlin2002}. 

Employing \eqref{i40} recursively, the global impedance matrix $\mathbf{Z}^{\textbf{K}}(r_k,r_0)$ for the   cylinder  between $r_0$ and $r_k$ is obtained with $3\times3$ components
\begin{equation}\label{i41}
\begin{aligned}
 \mathbf{Z}^{\textbf{K}}=
 \begin{pmatrix}
{{\mathbf Z}^{\textbf{K-1}}_{1}}-{\mathbf Z}^{\textbf{K-1}}_{2}\big({{\mathbf Z}}_{1}^{k} +{\mathbf Z}^{\textbf{K-1}}_{4} \big)^{-1} {\mathbf Z}^{\textbf{K-1}}_{3}
 & - {\mathbf Z}^{\textbf{K-1}}_{2} \big({{\mathbf Z}}_{1}^{k}  +{\mathbf Z}^{\textbf{K-1}}_{4} \big)^{-1} {{\mathbf Z}}_2^{k}
  \\ 
 - {{\mathbf Z}}_{3}^{k}   \big({{\mathbf Z}}_{1}^{k} +{\mathbf Z}^{\textbf{K-1}}_{4} \big)^{-1} {\mathbf Z}^{\textbf{K-1}}_{3}
  & {{\mathbf Z}}_{4}^{k}  - {{\mathbf Z}}_{3}^{k}   \big({{\mathbf Z}}_{1}^{k} +{\mathbf Z}^{\textbf{K-1}}_{4} \big)^{-1} {{\mathbf Z}}_2^{k} 
\end{pmatrix},
\end{aligned}
\end{equation}
where 
${\mathbf Z}^{\textbf{K-1}}_{i},\ (i=\overline{1,4})$  are the $3\times3$ sub-matrices of the    matrix ${\mathbf Z}^{\textbf{K-1}}(r_{k-1},r_{0})$ for $k-1$ layers, ${{\mathbf Z}}_{i}^{k},\ (i=\overline{1,4})$  are the $3\times3$  sub-matrices of the  matrix $\mathbf{Z}^{k}(r_{k},r_{k-1})$ for the $k\text{th}$ layer, defined by Eq.\ \eqref{i36}. 
The global impedance matrix for the $N$-layered cylinder is obtained by using Eq.\ \eqref{i41}  $(N-1)$ times.

  The main differences between the present results and those of \cite{Rokhlin2002} are, first that by construction the 
 local $\mathbf{Z}^{k}$ and global  $\mathbf{Z}^{\textbf{K}}$ two point impedance matrices are  Hermitian matrices.  Secondly, the present results are valid for cylindrically layered structures, as compared with those of  \cite{Rokhlin2002} which are for multilayered structures   in Cartesian coordinates.
 Despite the differences, we note that 
 the two point impedance matrix $\mathbf{Z}^{k}$ of Eq.\ \eqref{i35} and the global two point impedance matrix $\mathbf{Z}^{\textbf{K}}(r_k,r_0)$   are, apart from some sign differences,  similar to the stiffness matrix ${\bf K}^m$ and global stiffness matrix ${\bf K}^M$ of 
Rokhlin and Wang \cite{Rokhlin2002}.

\section{Stable solution technique for general anisotropy}\label{sec4}

In this section we propose and demonstrate a stable numerical scheme for solving for the conditional impedance matrix ${\bf z}(r)$ defined by \eqref{00} in the case of arbitrary radially dependent cylindrical anisotropy, i.e.\ density and elastic moduli are arbitrary functions of $r$: $\rho(r)$, $C_{ijkl}(r)$. 

\subsection{Stability issues}

\begin{figure}[h]
	\centering
		\includegraphics[width=5in]{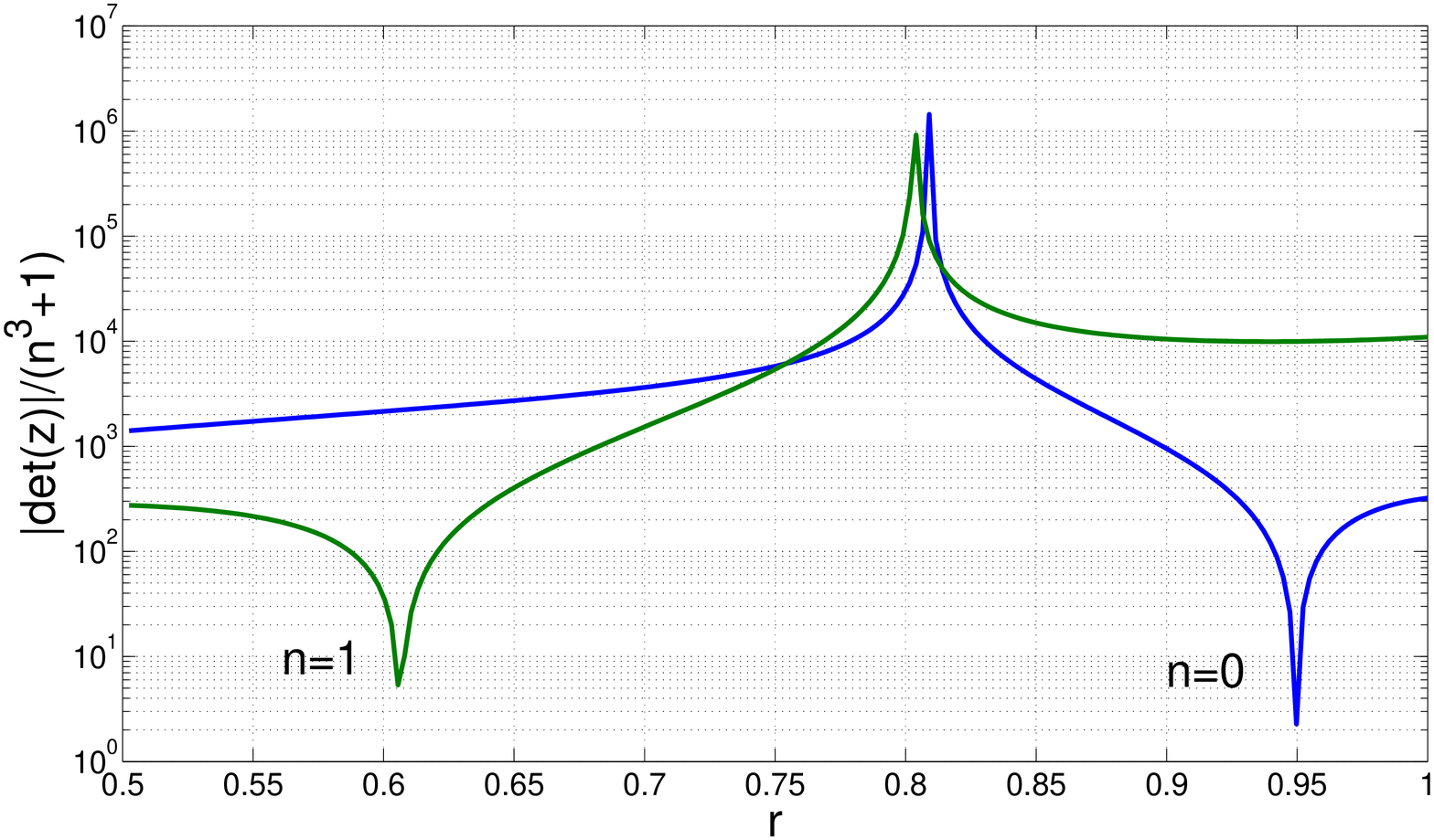}
		\caption{Solid, aluminium cylinder integrated with 200 evenly spaced steps, using the fourth order scheme of section \ref{fourthscheme}, from $r=0.5$ to $r=1.0$, with $k_z=0$ and $ka=10$.  Plotted is the determinant of the upper left $2$ $\times$ $2$ sub-matrix of the the $3$ $\times$ $3$ conditional impedance matrix normalized by $n^3+1$ vs. $r$.  Where Eq.\ \eqref{14z} was used for the initial impedance at $r=0.5$.}
	\label{fig1}
\end{figure}

Direct numerical integration of either  \eqref{2}  for the conditional impedance ${\bf z}(r)$ or \eqref{32}  for the matricant $\mathbf{M}(r,r_0)$  is not a feasible strategy.  The stiff nature of \eqref{32} leads to  exponentially growing instabilities  for $\mathbf{M}$.  These become unavoidable at large values of $n$ and/or $kr$ for the elastic problem.  
On the other hand, singularities and numeric instabilities may form when Eq.\ \eqref{2} is integrated, which is a well known issue for Riccati equations \cite{Schiff99}. 
The singularities (poles) of the impedance matrix occur at finite values of $r$  associated with traction-free modes for the given frequency.  One can, in principle, avoid the singularity by switching to the differential equation for the inverse of the impedance: the admittance $\mathbf{A} = {\bf z}^{-1}$ \cite[p. 136]{Bellman}.  The  admittance satisfies a Riccati differential equation, which follows from Eq.\ \eqref{2} as 
\begin{equation}\label{-3-99}
\frac{\dd \mathbf{A}}{\dd r} +\text{i} \mathbf{A}\mathbf{Q}_3 \mathbf{A} + \mathbf{A}\mathbf{Q}_4 
- \mathbf{Q}_1\mathbf{A} + \text{i}\mathbf{Q}_2 =0.
\end{equation}
However singularities again arise, this time   corresponding to rigid modes.  One could develop a numerical scheme that  switches back and forth  between integrating the impedance and admittance but since the locations of the singularities are not known in advance, and in practice they can be very close together, this does not offer a viable method. 
  The curves shown in Figure \ref{fig1} 
  exemplify the problem of singularities in the impedance.  
	The appearance of the peaks in Figure \ref{fig1} indicate values of $r$ beyond which accurate  numerical solution cannot be obtained regardless of the step size in the integration scheme.  	The inability of such standard methods to obtain correct results  was the motivation behind  the proposed solution technique discussed next.  

\subsection{M\"obius scheme}

Numerical  difficulties in  integrating the matrix Riccati equations are well known and have been studied extensively. 
A procedure mentioned earlier for avoiding singularities is a generalization of the idea of switching, where one undergoes a change of variables, which is closely related to invariant embedding 
methods, see e.g. \cite{Keller82}.  Here we follow a different approach, based on \cite{Schiff99},
which views the solution of the  Riccati equation   as a 
"Grassmanian flow" of $m$-dimensional subspaces (the impedance matrix) on a larger vector space of dimension $2m$.  The idea is to recast the equation for  ${\bf z} $ in the form 
of a forward marching scheme of step size $h$ based on \eqref{46}: 
\beq{47}
{\bf z} (r+h) = \text{i}\big( \mathbf{M}_{3}-\text{i}\mathbf{M}_{4}{\bf z} (r)  \big) \big( \mathbf{M}_{1}-\text{i}\mathbf{M}_{2}{\bf z} (r) \big)^{-1} , 
\eeq
where $\mathbf{M} = \mathbf{M}(r+h,r)$.    The key to the method is that $\mathbf{M}$ can always be calculated in a stable manner  for sufficiently small step size $h$.   This approach is one of a class of methods called M\"obius schemes, which by design are formulated  on the natural geometrical setting of the larger vector space, in this case that of $\mathbf{M}$.  Accordingly  M\"obius schemes are able to handle numerical instability and pass smoothly and accurately through  singularities \cite{Schiff99}.
The method therefore  combines both  the matricant and the impedance, each of which is unstable when solved in a global sense separately.  

\subsection{Approximations for $\mathbf{M}(r+h,r)$}

The M\"obius scheme shifts the problem to that of  finding approximations for $\mathbf{M}(r+h,r)$ accurate to some given order in $h$.   The  Peano series \cite{Pease} of cascading integrals for the matricant is formally  guaranteed to remain bounded for any $h$, but is numerically impractical.  Our objective is to develop 
 approximations for $\mathbf{M}(r+h,r) $ in the form 
\begin{equation}\label{22}
\mathbf{M}(r+h,r) = \mathbf{I}_{(2m)} + h \mathbf{M}^{(1)}(r)
+ \frac{h^2}{2!} \mathbf{M}^{(2)}(r) + \frac{h^3}{3!} \mathbf{M}^{(3)}(r)+\ldots , 
\end{equation}
where the terms $\mathbf{M}^{(2)}(r)$  do not require explicit integration schemes for their evaluation. 
Consider first the case of  $\mathbf{Q}$  sufficiently smooth.  Then  
 the identity 
\begin{equation}\label{21}
\mathbf{M}(r+h,r)= \mathbf{I}_{(2m)}+\int\limits_0^h \mathbf{Q}(r+s) \mathbf{M}(r+s,r) \dd s,  
\end{equation}
 may be written in the form of a series in powers of $h$ by using \eqref{22} for the left member
and a Taylor series evaluated at $r$ for $\mathbf{Q}$ in the integral, 
\begin{equation}\label{23}
\begin{aligned}
& h \mathbf{M}^{(1)} 
+ \frac{h^2}{2!} \mathbf{M}^{(2)} + \frac{h^3}{3!} \mathbf{M}^{(3)} 
+\ldots
= 
\int\limits_0^h \bigg(\big[\mathbf{Q}
+s \mathbf{Q}' + \frac{s^2}{2!}\mathbf{Q}''
\\ & \qquad \qquad + \frac{s^3}{3!}\mathbf{Q}''' +\ldots 
\big] \times 
\big[ \mathbf{I}_{(2m)} + s \mathbf{M}^{(1)} 
+ \frac{s^2}{2!} \mathbf{M}^{(2)} 
+\ldots 
\big]\bigg)\dd s ,  
\end{aligned}
\end{equation}
with the argument $r$ understood for all functions. 
Comparing equal orders of $h^k$ in \eqref{23} implies 
\begin{equation}\label{1212}
\begin{aligned}
 \mathbf{M}^{(1)} (r) &= \mathbf{Q} (r), \\
\mathbf{M}^{(2)} (r) &= \mathbf{Q}'(r) +\mathbf{Q}(r) \mathbf{M}^{(1)} (r),
\\
\mathbf{M}^{(3)} (r) &= \mathbf{Q}''(r) +2\mathbf{Q}'(r) \mathbf{M}^{(1)} (r)+\mathbf{Q}(r) \mathbf{M}^{(2)} (r),
\\
\mathbf{M}^{(4)} (r) &= 
\mathbf{Q}'''(r) +3\mathbf{Q}''(r) \mathbf{M}^{(1)} (r)+3\mathbf{Q}'(r) \mathbf{M}^{(2)} (r)
+\mathbf{Q}(r) \mathbf{M}^{(3)} (r),
\end{aligned}
\end{equation}
etc. The matricant may now be found by employing \eqref{22}. 
The series \eqref{1212}, while illustrative, is restricted to profiles that are analytically smooth functions of $r$.  It is not suitable for piece-wise constant or piece-wise smooth profiles, which are of practical interest.  Derivatives of the profile are therefore to be avoided if possible. In that sense, the approximation  formed from  Eqs.\ \eqref{22} and \eqref{1212}
is only valid to O$(h)$, and the iterative scheme \eqref{47} shares the same accuracy.

An expansion accurate to second order can be obtained by using 
a Taylor series expansion evaluated at the midpoint \cite{Schiff99}
\begin{equation}\label{2-134}
\begin{aligned}
\mathbf{Q}(r+s) \approx \mathbf{Q}(r+\frac{h}{2}) &+ (s-\frac{h}{2})\mathbf{Q}^{'}(r+\frac{h}{2}) + (s-\frac{h}{2})^2 \mathbf{Q}^{''} + \text{O}(s^3). 
\end{aligned}
\end{equation}
Substitution into \eqref{21} then  gives, using  \eqref{22}, 
\begin{equation}\label{59918}
\mathbf{M}^{(1)} = \mathbf{M}^{(2)}= \mathbf{Q}(r+\frac{h}{2}), \; \; \; \;
\mathbf{M}^{(3)}= \big(\mathbf{I}_{(2m)} + \frac{1}{2}\mathbf{Q}^{'}(r+\frac{h}{2})
\big)\mathbf{M}^{(1)}.
\end{equation}
This leads to an expansion up to $\text{O}(h^2)$ requiring only  $ \mathbf{Q}$ at a single position with no derivatives
\begin{equation}\label{25}
\text{TS } 2\text{nd}: \quad \mathbf{M}(r+h,r) = \mathbf{I}_{(2m)} + h \mathbf{Q} (r+\frac h2) + \frac{h^2}{2} \mathbf{Q}^{2}(r+\frac h2) +\text{O}(h^3) .
\end{equation}
The form of \eqref{25} suggests an alternative expression that is accurate to the same order
\begin{equation}\label{251}
\text{EXP } 2\text{nd}(a): \quad \mathbf{M}(r+h,r) = \exp\big(  h \mathbf{Q} (r+\frac h2) \big)
+\text{O}(h^3) .
\end{equation}
Approximations    \eqref{25} and   \eqref{251},  together with Eq.\ \eqref{47} each  yield a second order accurate M\"obius scheme.  EXP 2$^{nd}$ has the added feature that it is unimodular, and hence energy conserving.    
 Detailed comparisons are provided  in   section \ref{112-2}. 


\subsection{Lagrange interpolation expansions}
We now consider  Lagrange polynomial expansions for $\mathbf{Q}$ in Eq.\ \eqref{21}  in order to obtain higher order expressions without using derivatives of $\mathbf{Q}(r)$.   The Lagrange polynomial of order $n$ approximates 
$\mathbf{Q}(r+s)$ with the  expression \cite{Berrut}
\begin{equation}\label{2-11}
\mathbf{Q}(r+s) = \sum\limits_{j=0}^n \mathbf{Q} (r+x_j h) L_j\big(\frac sh\big), 
\qquad
L_j(x) = \prod_{l=0, l\ne j}^n \bigg( \frac{x -x_l}{x_j-x_l} \bigg),
\end{equation}
where $x_j \in [0,1]$, $j=0,1,\ldots , n$  are chosen points.  
Substituting into \eqref{23} 
and using the notation 
$\mathbf{Q}_{x_j} = \mathbf{Q} (r+x_j h)$ implies the sequence
\begin{equation}\label{2-114}
\begin{aligned}
\mathbf{M}^{(k)}   &= \bigg\{ \sum\limits_{j=0}^n  L_j^{(k)} \mathbf{Q}_{x_j} \bigg\}\, \mathbf{M}^{(k-1)} , \quad 
\mathbf{M}^{(0)}  \equiv \mathbf{I} , 
\\
\quad
 L_j^{(k)} &= k\int_0^1 
L_j(x) x^{k-1} \, \text{d} x , 
\end{aligned}
\qquad k=1,2,3\ldots .
\end{equation}
Note that $\sum\nolimits_{j=0}^n  L_j^{(k)} =1$. 
In the following subsections we derive expansions based on  \eqref{2-11} for $n=1$ and $n=3$.

\subsubsection{Two point approximation: Halves}
We approximate {\bf Q} with two points using \eqref{2-11} for $n=1$. In this case
the integrals $ L_j^{(k)} $ can be simplified with the result that 
\begin{equation}\label{2-22}
\mathbf{M}^{(k)}  = \bigg\{ \sum\limits_{j=0}^1  L_j \big(\frac{k}{k+1}\big)\mathbf{Q}_{x_j} \bigg\}\, \mathbf{M}^{(k-1)} , \quad 
\mathbf{M}^{(0)}  \equiv \mathbf{I} ,  
\quad 
 k=1,2,3\ldots .
\end{equation}
Taking equi-space points $x_0= \frac{1}{4}$ and $x_1 =  \frac{3}{4}$,  yields
\begin{equation}\label{2-33}
\text{LP } 2\text{nd}: \quad \mathbf{M}(r+h,r) = \mathbf{I}_{(2m)} + \frac h2 
\big(\mathbf{Q}_{\frac 14} +\mathbf{Q}_{\frac 34} \big)
+ \frac{h^2}{24} 
\big(\mathbf{Q}_{\frac 14} +5\mathbf{Q}_{\frac 34} \big)
\big(\mathbf{Q}_{\frac 14} +\mathbf{Q}_{\frac 34} \big)
+\text{O}(h^3) .
\end{equation}
This again gives a M\"obius scheme of  second order accuracy in $h$.  
It also suggests, by analogy with \eqref{251}, 
\begin{equation}\label{252}
\text{EXP } 2\text{nd}(b): \quad \mathbf{M}(r+h,r) = \exp\big(  \frac h2 \mathbf{Q}_{\frac 34}   \big)
\exp\big(  \frac h2 \mathbf{Q}_{\frac 14}   \big)
+\text{O}(h^3) .
\end{equation}
Note that  the expansions  \eqref{2-33} and \eqref{252}  only agree with one another to first order, O$(h)$.   

\subsubsection{Four point approximation: (Fourths)}\label{fourthscheme}
Taking four  evenly spaced  points to approximate $\mathbf{Q}$,  $x_j =  \frac{1}{8} + \frac{j}{4}$,   $j=0,1,2,3$, and using 
the  symbolic algebra program Maple, yields   
\begin{equation}\label{1599-13}
\text{LP } 4\text{th}: \quad
\begin{aligned}
\mathbf{M}^{(1)}=&\tfrac{13}{48}\mathbf{Q}_{\frac 18} + \tfrac{11}{48}\mathbf{Q}_{\frac 38} + \tfrac{11}{48}\mathbf{Q}_{\frac 58}+ \tfrac{13}{48}\mathbf{Q}_{\frac 78},
\\
\mathbf{M}^{(2)}=&\big(\tfrac{23}{720}\mathbf{Q}_{\frac 18} +\tfrac{67}{240}\mathbf{Q}_{\frac 38}+ \tfrac{43}{240}\mathbf{Q}_{\frac 58} 
+ \tfrac{367}{720}\mathbf{Q}_{\frac 78}\big)\mathbf{M}^{(1)},
\\
\mathbf{M}^{(3)}=&\big(-\tfrac{1}{48}\mathbf{Q}_{\frac 18} +\tfrac{19}{80}\mathbf{Q}_{\frac 38} + \tfrac{7}{80}\mathbf{Q}_{\frac 58} + \tfrac{167}{240}\mathbf{Q}_{\frac 78}\big)\mathbf{M}^{(2)},
\\
\mathbf{M}^{(4)}=&\big(-\tfrac{23}{560}\mathbf{Q}_{\frac 18} +\tfrac{389}{1680}\mathbf{Q}_{\frac 38}- \tfrac{67}{1680}\mathbf{Q}_{\frac 58} + \tfrac{1427}{1680}\mathbf{Q}_{\frac 78}\big) \mathbf{M}^{(3)}.
\end{aligned}
\end{equation}
Substitution of these terms into \eqref{22} gives $\mathbf{M}(r+h,r)$ up to fourth order accuracy.  Interestingly, when more points were taken to evaluate $\mathbf{Q}$ the  numerical accuracy was not found to improve. This was tried with  even spacings, using from five up to ten points.
Finally, by analogy with \eqref{252} we define 
\begin{equation}\label{253}
\text{EXP } 2\text{nd}(c): \quad \mathbf{M}(r+h,r) = 
\exp\big(  \tfrac h4 \mathbf{Q}_{\frac 78}   \big)
\exp\big(  \tfrac h4 \mathbf{Q}_{\frac 58}   \big)
\exp\big(  \tfrac h4 \mathbf{Q}_{\frac 38}   \big)
\exp\big(  \tfrac h4 \mathbf{Q}_{\frac 18}   \big)
+\text{O}(h^3) , 
\end{equation}
which, like \eqref{2-33} and \eqref{252}, is consistent with the four-term Lagrange interpolation scheme \eqref{1599-13} to first order. 
\subsection{Fourth order Magnus integrator scheme}
The Magnus integrator, created by Wilhelm  Magnus \cite{Magnus}, and further developed in \cite{Iserles99} with a convergence proof and recurrence relations, is a method to approximate a solution to \eqref{32} with 
\begin{equation}
\mathbf{M}(r+h,r)=e^{\mathbf{\Omega}} \mathbf{M}(r,r-h).
\end{equation}
Here we consider a fourth order Magnus integrator scheme similarly done in \cite{Lu2005} for the Helmholtz equation.  We use the following definitions to march a solution forward in $r$,
\begin{equation} \label{9948832}
\text{MG } 4\text{th}: \quad
\begin{aligned}
&\mathbf{M}(r+h,r)=e^{\mathbf{\Omega}} \mathbf{M}(r,r-h), 
\\
&\mathbf{\Omega} = \frac{h}{2}(\mathbf{Q}_{(1)} + \mathbf{Q}_{(2)}) + \frac{\sqrt{3}h^2}{12}(\mathbf{Q}_{(2)}\mathbf{Q}_{(1)} -\mathbf{Q}_{(1)}\mathbf{Q}_{(2)}),
\\
&\mathbf{Q}_{(1)}=\mathbf{Q}(r+h(\frac{1}{2} - \frac{\sqrt{3}}{6})), \quad \mathbf{Q}_{(2)}=\mathbf{Q}(r+h(\frac{1}{2} + \frac{\sqrt{3}}{6})).
\end{aligned}
\end{equation}
As a fourth order scheme the numerical precision of this method is very similar to that of the four point Lagrange polynomial approximation \eqref{1599-13}, which is seen in the examples of the following section.
\subsection{Numerical examples and convergence}\label{112-2}

\begin{figure}[h]
	\centering
		\includegraphics[width=5in]{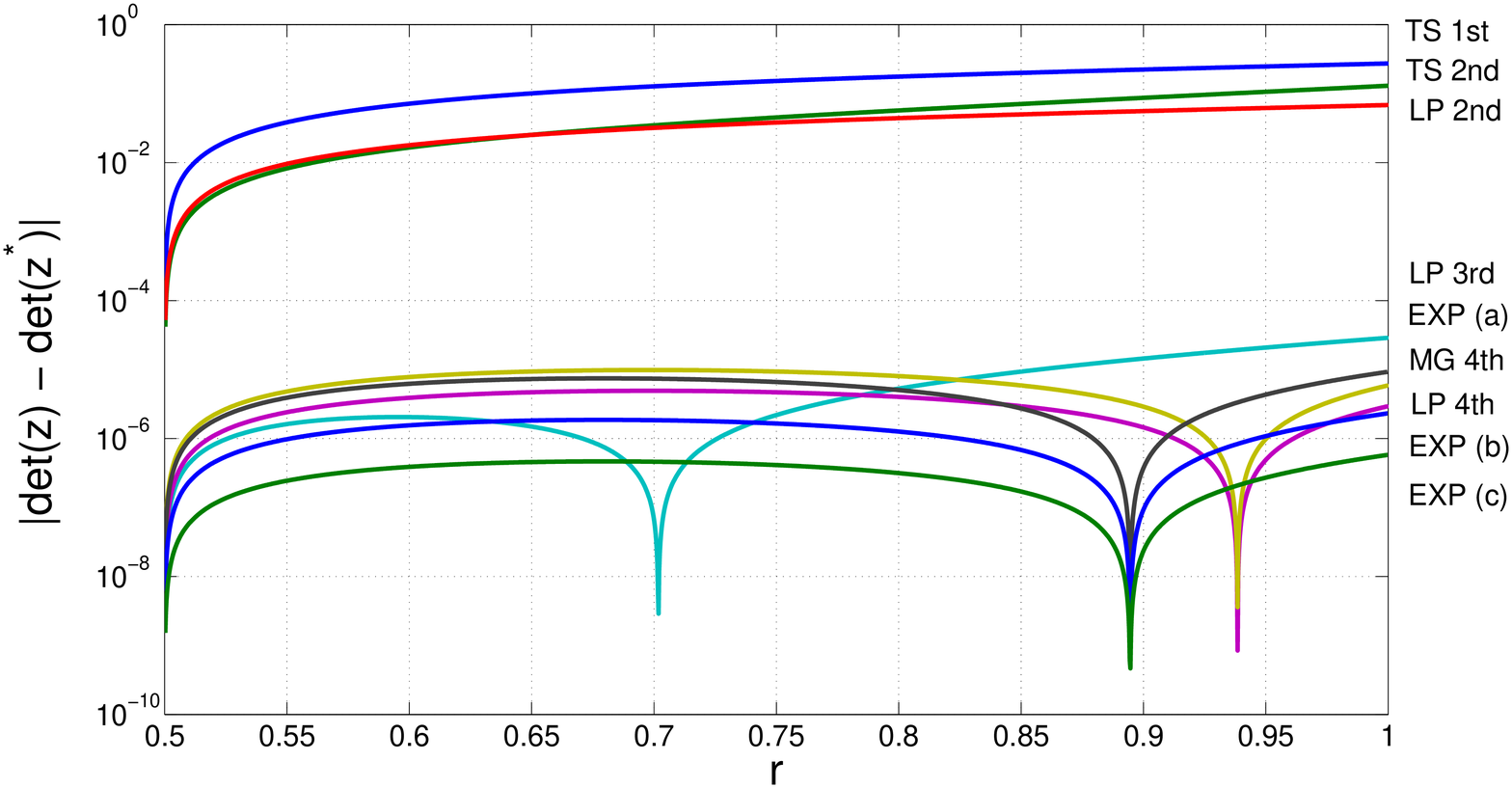}
		\caption{Solid, aluminium cylinder integrated with 2000 evenly spaced steps from $r=0.5$ to $r=1.0$, with $k_z=0$, $n=0$ and $ka=5$.  Plotted is the difference of the determinants of the upper left $2$ $\times$ $2$ sub-matrices of Eq.\ \eqref{14z} and those calculated from section \ref{sec4}.  As noted in section \ref{112-2} the right hand side refers to the type and order accuracy, they are listed top to bottom as worst to best at $r=1$.}
	\label{fig2}
\end{figure}
\begin{figure}[h]
	\centering
		\includegraphics[width=5in]{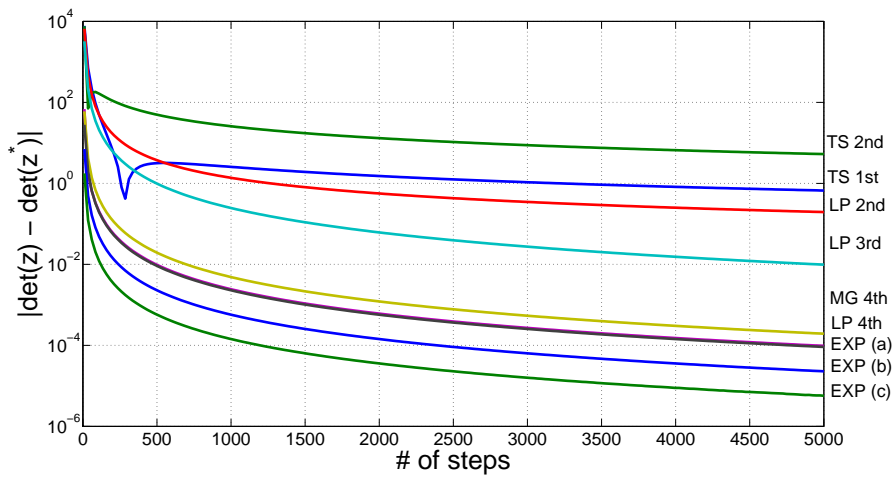}
		\caption{Solid, aluminium cylinder integrated from $r=0.5$ to $r=1.0$, with $k_z=0$, $n=0$ and $ka=25$.  Plotted is the difference of the determinants of the upper left $2$ $\times$ $2$ sub-matrices of Eq.\ \eqref{14z} and those calculated from section \ref{sec4} at $r=1$ vs. \# of steps. }
	\label{fig3}
\end{figure}

In order to illustrate the convergence property of the  different  expansions proposed (exponential, Magnus, Taylor series, Lagrange polynomials), we consider  a solid aluminium sample with properties $\rho=2.7 kg/m^3$, $E=70GPa$, $G=26GPa$ and radius of $r=1$ and normalized these properties with respect to water for which $\rho=1.0kg/m^3$ and speed of sound in water $c=1.470 km/s$.  The numerical values reported were computed by implementing the M\"obius scheme \eqref{47} and in the case of the Magnus integrator implementing \eqref{46}, starting at $r=0.5$ with initial condition given by the explicit solution \eqref{14z} from \cite{Norris10} and discussed in \S\ref{sec4}. In all examples we take $k_z=0$.  
Figure \ref{fig2} plots the difference of the determinant of the upper left $2$ $\times$ $2$ sub-matrix of the exact conditional impedance from \eqref{14z}, and that calculated by iterating 
\eqref{47} until  $r=1$ is reached.  The right hand side of Figure \ref{fig2} refers to the type
of approximation used: Taylor Series (TS), Lagrange Polynomial (LP), Exponential (EXP), Magnus (MG),  and to the  accuracy order.  Thus, LP 1st was calculated using Eq.\ \eqref{1212}, TS 2nd by \eqref{59918}, EXP (a) by \eqref{251}, EXP (b) by \eqref{252}, EXP (c) by \eqref{253}, LP 2nd by \eqref{2-33}, LP 4th by \eqref{1599-13}, MG 4th by \eqref{9948832}, and LP 3rd was calculated using a Lagrange Polynomial with points  $\{x_0, x_1,x_2\}$ $= \{\frac{1}{6}, \frac{1}{2},\frac{5}{6}\}$.  Interestingly, the three EXP methods (Eqs.\ \eqref{251}, \eqref{252} and \eqref{253}) gave similar results and were the best results for the fewest number of approximation points.
Figure \ref{fig3} plots the difference of the upper $2$ $\times$ $2$ sub-matrices at $r=1.0$ vs. the number of steps used in the iteration from $r=0.5$ to $r=1.0$.  As expected, the higher order schemes are more accurate and require fewer steps in the integration process to yield the same accuracy as a lower order scheme.

\section{Scattering Example}\label{sec5}
\begin{figure}[here]
\centering
	\includegraphics[width=5in]{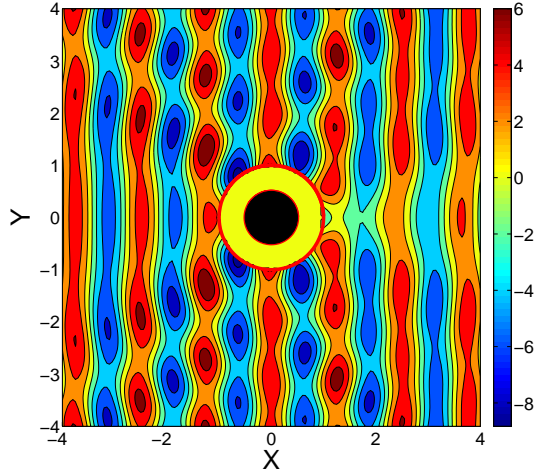}
	\caption{Plotted pressure field described by Eq.\ \eqref{151-15151} for an aluminium cylinder, integrated from $r=0.5$ to $r=1.0$ (area between the two red circles drawn) using the fourth order scheme, Eq.\ \eqref{1599-13}, with 500 steps.  The initial impedance at $r=0.5$ was found using Eq.\ \eqref{14z}. $ka=5$, $k_z=0$, and $\sigma_{tot}=2.468$.}
	\label{fig4}
\end{figure}
\begin{figure}[here] 
\centering
	\includegraphics[width=4in]{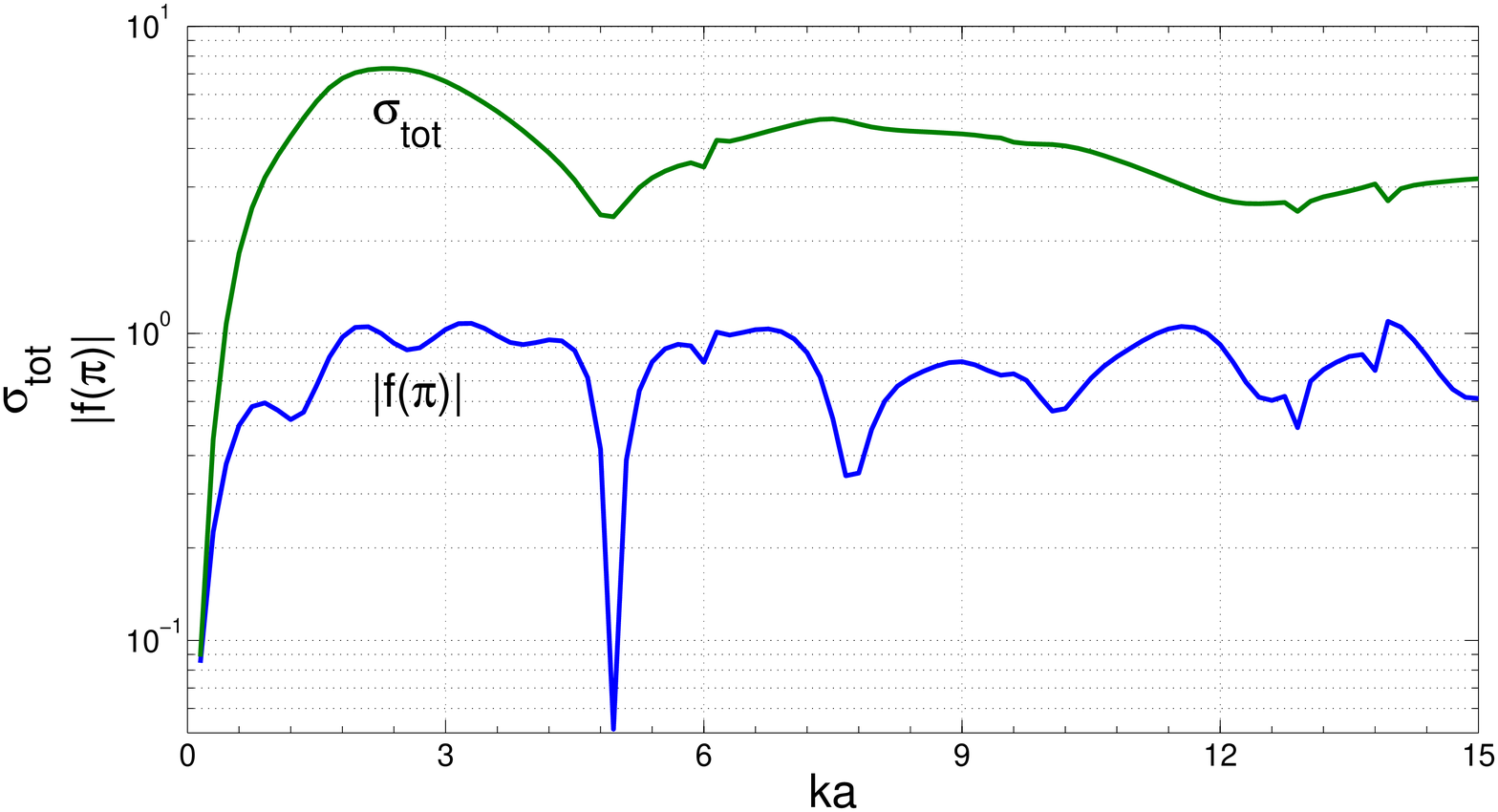}
	\caption{Total scattering cross section and backscattering amplitude plotted against non-dimensional frequency, $ka$.  Aluminum cylinder with the same properties as listed in figure \ref{fig4}, integrated using the fourth order scheme from $r=0.5$ to $r=1.0$ using 500 steps.  The initial impedance at $r=0.5$ was found using Eq.\ \eqref{14z}.}
	\label{fig5}
\end{figure}

In this section we explore the use of the impedance matrix by considering acoustic scattering from a solid aluminium cylinder immersed in water.  
Perpendicular plane wave incidence, i.e. $k_z=0$, in a uniform exterior fluid is considered with time harmonic dependence $\text{e}^{-\text{i}\omega t}$ henceforth omitted.  The total radial stress and displacement fields in the surrounding fluid are
\begin{equation}\label{151-15151}
\begin{aligned}
\sigma_{rr} &= - K k \sum\limits_{n=-\infty}^{\infty}  \text{i}^n\big(J_n(kr) + B_nH_n^{(1)}(kr)\big)\text{e}^{\text{i}n\theta},\\ u_r&=-\sum\limits_{n=-\infty}^{\infty}\text{i}^n\big(J_n^{'}(kr) + B_nH_n^{(1)'}(kr)\big)\text{e}^{\text{i}n\theta},
\end{aligned}
\end{equation}
where $r$ is the radial coordinate, $K$ is the bulk modulus, $k$ is the wave number,  $H_n^{(1)}$ is the Hankel function of the first kind, and the coefficients $B_n$ are to be determined \cite{MorseI}. We use the definition of the conditional impedance matrix, noted in Eq.\ \eqref{00}, and write this statement for the innermost radial coordinate at $r=b$, for which we find the initial impedance matrix from Eq.\ \eqref{14z}, and for the outer surface at $r=a$ where
\begin{equation}\label{51-151}
\mathbf{V}(b) = -\text{i}{\bf z}_1\mathbf{U}(b), \; \; \; \; \mathbf{V}(a) = -\text{i}{\bf z}_2\mathbf{U}(a).
\end{equation}
The conditional impedance matrix, ${\bf z}_2 = {\bf z}(a)$, is found from the integration technique outlined in section \ref{sec4}, or for transversely isotropic materials may be found directly using Eq.\ \eqref{14z}.  Considering acoustic fluid in the exterior we write Eq.\ \eqref{51-151} for $r=a$ in detail for which the shear stress, $\sigma_{r\theta}$, must be zero with
\begin{equation}\label{41-151}
\text{i}a\begin{pmatrix} \sigma_{rr} \\ 0 \end{pmatrix} = -\text{i}{\bf z}_2\begin{pmatrix} u_r \\ u_{\theta} \end{pmatrix} = -\text{i}\begin{pmatrix} p_1 & q_1 \\ p_2 & q_2 \end{pmatrix}\begin{pmatrix} u_r \\ u_{\theta} \end{pmatrix}.
\end{equation}
Eliminating $u_{\theta}$ using the second row of Eq.\ \eqref{41-151} implies
\begin{equation}\label{151-1636}
\text{i}a \frac{\sigma_{rr}}{u_r}= \frac{\text{i}}{ q_2} (q_1p_2 - q_2 p_1) \equiv -\text{i} z_0.
\end{equation}
Using \eqref{151-15151} and equating with \eqref{151-1636} we find the scattering coefficient
\begin{equation}
B_n = - \frac{K k a J_n(ka) + z_0J_n^{'}(ka)}{K k a H_n^{(1)}(ka) +z_0H_n^{(1)'}(ka)}.
\end{equation}
Numerical simulation   was conducted for  a solid aluminium cylinder with properties normalized with respect to water and with the total pressure field   illustrated in Figure \ref{fig4}.
Figure \ref{fig5}  shows both the total scattering cross-section $\sigma_{tot}$ and the 
back-scattering amplitude $f(\pi)$, where 
the far field form function, $f(\theta)$ is 
\begin{equation}
f(\theta)=\frac{2}{\sqrt{k}}\sum_{n=0}^{\infty}  \text{i}^{2n-1}\epsilon_n B_n  \cos n\theta,
\end{equation}
where $\epsilon_0 = 1$ and $\epsilon_m=2$ for $m>0$.  The total scattering cross section is then  
\begin{equation}
\sigma_{tot} = \frac{4\pi}{ka} \text{Imag}(f(0)).
\end{equation}
Figure \ref{fig5} closely matches the behavior of a similar analysis conducted in \cite{Flax}.  


\section{Conclusion}  \label{sec6}
Two computationally stable methods were considered to calculate the impedance matrix. Seeking solutions of 3D elasticity in the form of time-harmonic cylindrical waves, a matrix  Riccati equation for the impedance matrix was formulated. Direct integration of the Riccati equation is numerically unstable, it leads to exponentially growing instabilities, which  is inescapable at high frequency, large values of  $n$, and layer thickness. We integrated the Riccati equation for the impedance matrix which has singularities along the real radial coordinate associated with traction free modes. To avoid instabilities we developed a new stable numerical scheme for cylindrically anisotropic structures that passes through these singularities by combining the impedance and  matricant.  This scheme evaluates the impedance matrix for continuous systems by integrating
the Riccati equation over the thickness of each layer.
 Different expansion methods were considered  and compared, it was noted that matrix exponential approximations yielded the best results. 

An alternative method was developed to obtain the global impedance matrix for anisotropic, cylindrically layered media using the impedance matrix for each layer. The recursive formula to calculate the total two point impedance was derived. The impedance matrix method was applied to obtain the total surface impedance matrix calculated recursively, layer by layer by employing the recursive formula $N-1$ times for an $N$ layered system.  In the case of more complex inhomogeneity
 and a large number of cylindrical layers,  the recursive
algorithm and the alternative integration technique are far superior to methods involving finite element analysis and can be performed with completely anisotropic materials.


Application of the impedance matrix simplifies the formulation of various scattering and boundary value problems for cylindrical structures. The impedance matrix can be applied to solve various acoustic and elastic scattering problems  for arbitrarily layered cylindrical shells and solids.  As an example acoustic scattering from a solid aluminium cylinder immersed in water was considered using the impedance matrix and compared with the literature. Plots of the total pressure field, form function and total scattering cross section agree with previously published results.

\appendix
\section{Cylindrically anisotropic media}\label{appA}
The equilibrium equations of linear elastodynamics in cylindrical coordinates are  \cite{Shuvalov03}
\begin{equation} \label{-1-1}
(r{\bf t}_r)_{,r} + {\bf t}_{\theta,\theta} + {\bf K}{\bf t}_{\theta} + r{\bf t}_{z,z} = r \rho \ddot{{\bf u}},
\end{equation}
where 
 ${\bf u} = (u_r ,\, u_{\theta} ,\, u_z )^T$ is the displacement vector, $\rho$ is the mass density, 
 ${\bf t}_r$, ${\bf t}_{\theta}$, ${\bf t}_z$, 
are the traction vectors,  ${\bf K}$ is the $3$ $\times$ $3$  matrix with $K_{12}=-1$, $K_{21}=1$ and zero otherwise,  $T$ denotes transpose and
a  comma suffix denotes partial differentiation.  Using the index notation $(r, \theta, z) \leftrightarrow (1, 2, 3)$  the stress strain relationship is
\begin{equation}\label{-1-6}
\sigma_{ij} = C_{ijkl}\varepsilon_{kl} \ \ \big( =({\bf t}_i)_j \big) ,
\end{equation}
where the elastic stiffness tensor elements have the usual symmetries 
$C_{ijkl}=C_{jikl}=C_{klij}$, 
and the  notation of  summation over repeated indices is assumed.
Combining the displacement strain equations in cylindrical coordinates with Eq.\ \eqref{-1-6} the traction vectors are \cite{Shuvalov03}
\begin{equation}\label{-1-9}
\begin{pmatrix} {\bf t}_r \\ {\bf t}_\theta \\ {\bf t}_z \end{pmatrix} = \begin{pmatrix} \hat{{\bf Q}} & {\bf R} & {\bf P} \\ {\bf R}^T & \hat{{\bf T}} & {\bf S} \\ {\bf P}^T & {\bf S}^T & \hat{{\bf M}} \end{pmatrix} \begin{pmatrix} {\bf u}_{,r} \\ r^{-1}({\bf u}_{,\theta} + {\bf K}{\bf u}) \\ {\bf u}_{,z} \end{pmatrix} .
\end{equation}
  The form of the various matrices in \eqref{-1-9} are, using Voigt notation for the moduli,
\begin{align}\label{-1-10}
\hat{{\bf Q}} &= \begin{pmatrix} C_{11} & C_{16} & C_{15} \\ C_{16} & C_{66} & C_{56} \\ C_{15} & C_{56} & C_{55} \end{pmatrix}, \;  
\hat{{\bf T}} = \begin{pmatrix} C_{66} & C_{26} & C_{46} \\ C_{26} & C_{22} & C_{24} \\ C_{46} & C_{24} & C_{44} \end{pmatrix}, \;
 \hat{{\bf M}} = \begin{pmatrix} C_{55} & C_{45} & C_{35} \\ C_{45} & C_{44} & C_{34} \\ C_{35} & C_{34} & C_{33} \end{pmatrix}, 
\\
 {\bf R} &= \begin{pmatrix} C_{16} & C_{12} & C_{14} \\ C_{66} & C_{26} & C_{46} \\ C_{56} & C_{25} & C_{45} \end{pmatrix}, \;
 {\bf P} = \begin{pmatrix} C_{15} & C_{14} & C_{13} \\ C_{56} & C_{46} & C_{36} \\ C_{55} & C_{45} & C_{35} \end{pmatrix}, \; 
 {\bf S} = \begin{pmatrix} C_{56} & C_{46} & C_{36} \\ C_{25} & C_{24} & C_{23} \\ C_{45} & C_{44} & C_{34} \end{pmatrix}.
\notag
\end{align}

We consider  cylindrically anisotropic media, illustrated in Figure \ref{fig0}, for which  the density, $\rho$, and elasticity tensor, ${\bf C}$,  depend on the radial coordinate $r$.  Solutions 
are sought in the form of time-harmonic cylindrical waves where the displacement and radial traction vectors are of the form 
\begin{equation}\label{-1-11}
{\bf u} = \mathbf{U}^{(n)}(r)\text{e}^{\text{i}(n\theta + k_z z- \omega t)}, \quad
\text{i} r {\bf t}_r = \mathbf{V}^{(n)}(r)\text{e}^{\text{i}(n\theta + k_z z - \omega t)},
\end{equation}
where $\omega$ is the frequency, $k_z$ is the axial wave number,  and $n$ $=$ $0$, $1$, $2$, $\ldots$ is the circumferential number.  Plugging \eqref{-1-11} into \eqref{-1-1} and \eqref{-1-9} yields the  state space representation \cite{Shuvalov03}
for the state vector $\boldsymbol{\eta}^{(n)}$ consisting of the displacement and radial traction  vectors, 
\begin{equation}\label{-1-12}
\frac{\dd}{\dd r} \boldsymbol{\eta}^{(n)}(r) = \frac{\text{i}}{r} {\bf G}(r)\boldsymbol{\eta}^{(n)}(r) 
\quad \text{with }\ \ 
\boldsymbol{\eta}^{(n)}(r) = \begin{pmatrix} \mathbf{U}^{(n)}(r) \\ \mathbf{V}^{(n)}(r) \end{pmatrix}.
\end{equation}
The superscript $(n)$ is omitted for the remainder of the paper.  The $6$ $\times$ $6$ system matrix is  \cite{Shuvalov03} 
\begin{equation}\label{-1-14}
\text{i}{\bf G} = \begin{pmatrix} {\bf g}^{\{1\}} & \text{i}{\bf g}^{\{2\}} \\ \text{i}{\bf g}^{\{3\}} & -{\bf g}^{\{1\}+} \end{pmatrix},
\end{equation}
where all terms depend on the radial coordinate, $r$, and superscript $+$ denotes the adjoint or conjugate transpose of a matrix.  The $3$ $\times$ $3$ matrices in \eqref{-1-14} are
\begin{equation}\label{-1-16}
\begin{aligned}
{\bf g}^{\{1\}}&=-\hat{{\bf Q}}^{-1}\tilde{{\bf R}} - \text{i}k_zr\hat{{\bf Q}}^{-1}{\bf P}, \quad 
{\bf g}^{\{2\}}=  {\bf g}^{\{2\}T}= -\hat{{\bf Q}}^{-1}, \\
{\bf g}^{\{3\}} &= {\bf g}^{\{3\}+}
=\tilde{{\bf T}}-\tilde{{\bf R}}^{+}\hat{{\bf Q}}^{-1}\tilde{{\bf R}}
+ \text{i}k_zr \big[ {\bf P}^T\hat{{\bf Q}}^{-1}\tilde{{\bf R}}-\tilde{{\bf S}}-({\bf P}^T\hat{{\bf Q}}^{-1}\tilde{{\bf R}}-\tilde{{\bf S}})^+   \big]  
\\ &\qquad \qquad \qquad 
 + r^2\big[  k_z^2(\hat{{\bf M}} - {\bf P}^T\hat{{\bf Q}}^{-1}{\bf P}) - \rho \omega^2 {\bf I} \big],
\end{aligned}
\end{equation}
${\bf I}$ is the $3$ $\times$ $3$ identity matrix, $^+$ indicates Hermitian transpose, and 
\begin{equation}\label{-1-18}
\tilde{{\bf R}}={\bf R}\boldsymbol{\kappa}, \; \; 
\tilde{{\bf S}}=\boldsymbol{\kappa}{\bf S}, \; \;\tilde{{\bf T}}=\tilde{{\bf T}}^{+} = \boldsymbol{\kappa}^{+}\hat{{\bf T}}\boldsymbol{\kappa},  \quad\text{with }\  
\boldsymbol{\kappa} = {\bf K} + in{\bf I} \ \big(=-\boldsymbol{\kappa}^{+}\big).
\end{equation}
The system matrix, ${\bf G}$ has the important symmetry \cite{Shuvalov03} which follows from  the form of \eqref{-1-14}, 
\begin{equation}\label{-1-19}
{\bf G} = {\bf T}{\bf G}^{+}{\bf T}, \; \; \text{with} \; \; {\bf T} = \begin{pmatrix} \boldsymbol{0} & {\bf I} \\ {\bf I} & \boldsymbol{0}\end{pmatrix}.
\end{equation} 
The problem is now reduced to finding a solution to Eq.\ \eqref{-1-12}$_1$ subject to appropriate boundary conditions.  Next we introduce  the matricant and impedance matrices based on solutions  of \eqref{-1-12}$_1$.

\section{Impedance for uniform transverse isotropy} \label{appB}

We consider transversely isotropic solids with the symmetry axis in the $z-$direction. 
The displacement vector may be decomposed using Buchwald's scalar potentials \cite{Buchwald}, the  functions $\varphi, \, \chi$ and $\psi$,  
\beq{R23}
{\bf u} =\boldsymbol{\nabla} \varphi + \boldsymbol{\nabla} \times (\chi {\bf e}_z)    + \bigg( \frac{\partial \psi}{\partial z} - \frac{\partial \varphi}{\partial z} \bigg) {\bf e}_z.
 \eeq
The general solution of the  equilibrium  equations for transverse   isotropy are of the form $\{ \varphi,  \chi ,\psi \} = \{ \bar\varphi,  \bar\chi ,\bar\psi \}
\text{e}^{\text{i}(n \theta+k_z z- \omega t)}$ where  
\beq{R35Xb}
\begin{small}
\begin{aligned}
\bar\varphi & =   R_{on}^{\,l}\, \frac{1}{k_1} f_n^{\,l} (k_1 r)+ \frac{1}{k_2}  S_{on}^{\,l} \,f_n^{\,l} (k_2 r) ,
\\
\bar\psi &=  \frac{\kappa_1}{k_1}\, R_{on}^{\,l}\, f_n^{\,l} (k_1 r)+ S_{on}^{\,l} \frac{\kappa_2}{k_2} \,f_n^{\,l} (k_2 r)  ,  
\end{aligned}
\quad 
\bar\chi  =  -   \,T_{on}^{\,l}\,\frac{1}{k_3} f_n^{\,l} (k_3 r)  , 
\end{small}
\eeq
and  $R_{on}^{\,l},\,S_{on}^{\,l},\, T_{on}^{\,l}$ are unknown coefficients, $f_n^{\,l}(x)$ are cylindrical functions: $f^{\,1}_n(x) = J_n(x)$  for solutions that are regular at $r=0$, $f^{\,2}_n(x) = Y_n(x)$ for real valued irregular solutions at $r=0$, $f^{\,3}_n(x) = H_n^{(1)}(x)$ for outgoing (radiating) solutions, $f^{\,4}_n(x) = H_n^{(2)}(x)$ for ingoing solutions, where $ J_n(x)$ - Bessel function of the first kind; $Y_n(x)$ - Bessel function of the second kind;  $ H_n^{(1,2)}(x)$ - Hankel functions of the first and second kind. The displacement field can be represented  as a linear combination of any two  of the four types of cylindrical functions $f^{\,l}_n(x),\, (l=\overline{1,4})$. 
The wavenumbers $k_1$, $k_2$, $k_3$, and non-dimensional numbers $\kappa_1$, $\kappa_2 $ are given by 
\beq{R32}
\begin{small}
\begin{aligned}
k_{1,\,2}^2 & =\frac{ a \mp \,\sqrt{a^2 - b}}{2 \,c_{11} \, c_{44}}, \quad k_3^2 = \frac{\rho \, \omega ^2 -c_{44} k_z^2}{c_{66}},
\quad \kappa_i= \frac{c_{66}\,k_3^2 - c_{11}\, k_i^2  }{\big(c_{13} + c_{44} \big) k_z},\quad (i=1,2),
\\
a \, &=\,\big(
c_{11}
 +\, c_{44} \big)\rho \, \omega ^2 + \big( c_{13}^2+2 c_{13}c_{44} - c_{11} \, c_{33}  \big)k_z^2,  
\\
b\, &=\,4 c_{11} \, c_{44}\, \big( \rho \omega^2 -c_{33} \, k_z^2 \big)  \big( \rho \omega ^2 -\, c_{44} \, k_z^2 \big).
\end{aligned}
\end{small}
\eeq
For isotropic material wavenumbers $k_i, \,\kappa_i$ reduce to 
$k_{1}^2 = \omega ^2{\rho }/({\lambda+2\mu})  - k_z^2$, $k_{2}^2= k_{3}^2=  \omega ^2{\rho }/{\mu} - k_z^2$, $
\kappa_1 = 1$, $  \kappa_2 = - {k_2^2}/{k_z^2}$.

The displacement and traction vectors  
${\bf U}$ and ${\bf V}$ of \eqref{-1-11} are obtained in matrix form for each $n$ as 
\beq{R17}\begin{small}
{\bf U} (r) 
= 
\sum_l
{\bf X}^l(r) 
{\bf w}^l ,
\quad
{\bf V} (r) 
=\sum_l
{\bf Y}^l(r) 
{\bf w}^l ,
\quad
{\bf w}^l = 
\begin{pmatrix} 
R_{on}^{\,l}
\\
S_{on}^{\,l}
\\
T_{on}^{\,l}
\end{pmatrix} ,
 \end{small}
\eeq
where the summation on $l$ is over any two of the possible  $l=\overline{1,4}$,  and 
\begin{align}\label{14U}
{\bf X}^l(r) &=
\begin{bmatrix} 
{f^{\,l}_{n}}'(k_{1}r) & {f^{\,l}_{n}}'(k_{2}r) &  - \frac{\text{i} n }{k_{3}r}f^l_{n}(k_{3}r)
\\
\,\frac{\text{i}n }{k_{1}r} f^l_{n}(k_{1}r)&   \frac{\text{i} n }{k_{2}r} f^l_{n}(k_{2}r) &  {f^l_{n}}'(k_{3}r)
\\
 \frac{ \text{i} \kappa_1}{k_{1}}f^l_{n}(k_{1}r) &    \frac{\text{i} \kappa_2}{k_{2}} f^l_{n}(k_{2}r)  & 0
 \end{bmatrix} , 
\\
 {\bf Y}^l(r)&= - \text{i} {\bf z}^l(r) {\bf X}^l(r) , 
\end{align}
and ${\bf z}^l$, $l=\overline{1,4}$, follows from \cite{Norris10}:
\begin{equation}\label{14z}\begin{small}
\begin{aligned}
{\bf z}^l(r) &= \begin{pmatrix} 
2 c_{66} & \text{i}n2c_{66} & \text{i} k_z r c_{44}
\\
-\text{i}n 2c_{66} & 2c_{66} & 0
\\
- \text{i} k_z r c_{44} & 0 & Z_z
 \end{pmatrix}
\\&
+ c_0
\begin{pmatrix} 
\xi_3 (y_1-y_2) & \text{i}n(y_1-y_2) & \text{i} \xi_3 (\xi_1-\xi_2)
\\
- \text{i}n (y_1-y_2) & \xi_2y_1-\xi_1 y_2 & n  (\xi_1-\xi_2)
\\
-\text{i} \xi_3 (\xi_1-\xi_2) & n (\xi_1-\xi_2) &0
 \end{pmatrix},
\end{aligned}
\end{small}
\end{equation}
\beq{14z1}
\begin{aligned}
Z_z&= c_{44}\bigg( \frac{n^2(\xi_1 y_1-\xi_2y_2)-\xi_1 \xi_2 \xi_3(y_1-y_2)}{\xi_3(\xi_2y_1-\xi_1 y_2)-n^2(y_1-y_2)}  \bigg), \quad y_i= \kappa_i r \quad (i=1,2),
\\
c_0&= \frac{c_{66}k_3^2 r^2}{\xi_3(\xi_2y_1-\xi_1 y_2)-n^2(y_1-y_2)} , \quad \xi_j=k_j r \frac{{f^l_{n}}'(k_{j}r)}{f^l_{n}(k_{j}r)} \quad (j=1,2,3).
\end{aligned}
\eeq
The formula for ${\bf X}^l$  follows by substituting the  potentials \eqref{R35Xb} into Eq.\ \eqref{R23}.
The derivation of the matrix ${\bf z}^l(r)$ can be found in \cite{Norris10}.  Note that 
${\bf z}^1(r)$ $(l\equiv 1)$ is the exact form of the conditional impedance of a solid cylinder, i.e.\ with material at $r=0$ and hence bounded displacements there \cite{Norris10}. 

The explicit form of the two point impedance matrix (see Eq.\  \eqref{i35}) of a given  transversely isotropic layer is 
\beq{i36}\begin{small}
\mathbf{Z}^{k} (r_{k},r_{k-1}) =\begin{pmatrix}
{\mathbf{Z}}_{1}^{k}  & {\mathbf{Z}}_{2}^{k} \\ 
{\mathbf{Z}}_{3}^{k} & {\mathbf{Z}}_{4}^{k}
\end{pmatrix}=   
\begin{bmatrix} 
-{\bf Y}^{1}(r_{k-1}) & -{\bf Y}^{3}(r_{k-1})
\\
 {\bf Y}^{1}(r_{k}) &  {\bf Y}^{3}(r_{k})
\end{bmatrix}
\begin{bmatrix} 
{\bf X}^{1}(r_{k-1}) & {\bf X}^{3}(r_{k-1})
\\
{\bf X}^{1}(r_{k})\  & {\bf X}^{3}(r_{k}) \ 
\end{bmatrix}^{-1}.
\end{small}\eeq
Equation \eqref{i36}, which defines the impedance matrix $\mathbf{Z}$,  is similar to Eq.\ (7)  of  \cite{Rokhlin2002} (for  the stiffness matrix $\mathbf{K}$), and the first and the second matrices on the right hand side of Eq.\ \eqref{i36} are similar to the matrices ${\bf E}_m^{\sigma}$ and $({\bf E}_m^u)^{-1}$  in \cite[Eqs.\ (5) and (3)]{Rokhlin2002}.  One reason why we prefer to use the impedance matrix $\mathbf{Z}$ rather  than the stiffness matrix as in \cite{Rokhlin2002} is that the impedance is always Hermitian:  $\mathbf{Z}=\mathbf{Z}^{+}$.

\section*{Acknowledgments}  This work was supported by the National Science Foundation and by the Office of Naval Research.

\bibliographystyle{unsrt}

\end{document}